# Effective Correlates of Motor Imagery Performance based on Default Mode Network in Resting-State

Jae-Geun Yoon[1], Minji Lee[1]
[1]Department of Brain and Cognitive Engineering, Korea University, Seoul, Korea
jg_yoon@korea.ac.kr, minjilee@korea.ac.kr

*Abstract*—Motor imagery based brain-computer interfaces (MI-BCIs) allow the control of devices and communication by imagining different muscle movements. However, most studies have reported a problem of "BCI-illiteracy" that does not have enough performance to use MI-BCI. Therefore, understanding subjects with poor performance and finding the cause of performance variation is still an important challenge. In this study, we proposed predictors of MI performance using effective connectivity in resting-state EEG. As a result, the high and low MI performance groups had a significant difference as 23% MI performance difference. We also found that connection from right lateral parietal to left lateral parietal in resting-state EEG was correlated significantly with MI performance ($r = -0.37$). These findings could help to understand BCI-illiteracy and to consider alternatives that are appropriate for the subject.

## I. INTRODUCTION

Brain-computer interface (BCI) provides a system that can control an external electronic device based on human intentions by using electrical signals generated from the brain [1-4]. This has been of great interest as an emerging technology in medical or industrial applications such as treatment and rehabilitation as a tool without direct movements such as muscle activity [5-6]. Specifically, motor imagery-based BCI (MI-BCI) is used to distinguish the intention of the user based on the imagination of the actual muscle movement [7-9]. Unfortunately, MI-BCI still have a big obstacle to use in the public market. For MI based on voluntary brain signals, performance has been observed to vary across different inter-subject or inter-experiment. In addition, problems have been reported in which 15~30% of users cannot control the MI-BCI. This phenomenon is called "BCI-illiteracy" [10-11]. Therefore, it is important to understand the cause for the individual differences in MI-BCI performance and find reliable biomarkers to predict individual MI-BCI performance. The proposition of new predictors understands the common causes of inefficient MI-BCI subjects and identification of subjects with poor performance can save time and resources. Studies dealing with these problems may be suitable training alternatives for subjects who exhibit poor performance on task.

This work was supported by Institute for Information & Communications Technology Promotion (IITP) grant funded by the Korea government (No. 2017-0-00451, Development of BCI based Brain and Cognitive Computing Technology for Recognizing User's Intentions using Deep Learning).

Many related studies have suggested neurophysiological or psychological predictors for the performance of MI-BCI. In the case of psychological predictors, questionnaires before the experiment are used to examine the relationship with MI performances. For example, user frustration showed a significant correlation with MI performance [12]. Jeunet et al. [13] were investigated the relationship between the BCI performance and personality including cognitive profile. According to Ahn et al. [14], the user's self-prediction showed the possibility to predict MI performance ($r = 0.54$). On the other hand, related studies suggesting neurophysiological predictors have demonstrated the relationship with MI using the features of electroencephalography (EEG) signals. In Blankertz et al. [15], sensory motor rhythms (SMRs) of 2-min resting-state EEG from 80 subjects were proposed as a predictor and showed a significant correlation with MI performance ($r = 0.53$). In addition, the spectral entropy of the C3 channel in resting EEG was highly correlated with the performance of SMR-BCI ($r = 0.65$). Ahn et al. [16] showed the possibility that the theta and alpha powers of resting-state EEG predicted MI performance. Specifically, they showed that high theta and low alpha waves in resting-state EEG were noticeable in the BCI-illiteracy ($r = 0.59$). The above studies were proposed predictors to reveal the underlying neural mechanisms, but there is still difficulty in understanding the common features of many subjects with BCI-illiteracy [17].

Development of neuroimaging techniques, such as EEG, functional magnetic resonance imaging (fMRI), and structural MRI (sMRI) has improved the understanding of mechanisms related to BCI [18-20]. According to Zhang et al. [17], they demonstrated that an efficient resting EEG network facilitates MI performance. They also observed that features of resting-state EEG based on functional connectivity correlated significantly with the MI performance. Interestingly, Zhang et al. [21] have also been shown to improve the performance of subjects with BCI-illiteracy using features of brain networks. In general, features of the brain network can be described as functional connectivity or effective connectivity. Functional connectivity is usually inferred based on correlations among measurements of neuronal activity but it does not provide any directional information [22]. On the other hand, effective connectivity constitutes a directional brain network and explains the causal relationship between brain areas [23-25]. In this sense, effective connectivity analysis summarizes the scientific process about how the observed data occurred. Through this difference, effective connectivity can be used to better understand brain phenomena when performing tasks [26-

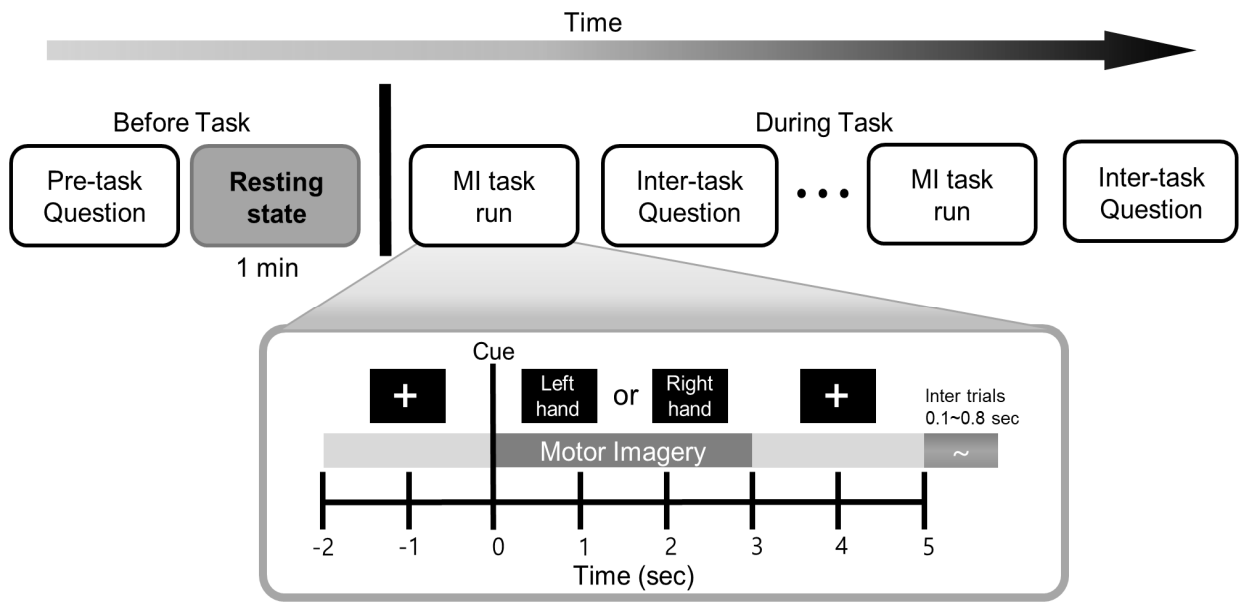

Figure 1. Experiment design. Before the MI task, 1 min resting-state EEG with eyes open was recorded. MI task for left and right hand movement was performed. Each run of MI task include 100 or 120 trials.

27]. In the case of dynamic causal modeling (DCM), it can build the new connectivity of the new brain network and quantify the effective connectivity to the brain region [28]. Kim et al. [29] investigated the relationship between motor-relation and frontal areas using DCM of MI and motor execution (ME). As a result, they demonstrated that the frontal area was involved in the MI and ME tasks when using DCM in the region of interest (ROI) of MI-relation area and frontal area. Studies for fundamental mechanisms needed for performing MI tasks can help improve the performance of subjects with poor performance. Therefore, this study was conducted to find a mechanism related to MI using the change of effective connectivity in resting-state EEG.

In this study, we investigated the relationship between effective connectivity in resting-state EEG and MI performance. Specifically, the subjects were divided as high and low groups according to MI classification accuracy and analyzed for common differences between two groups of MI performance. We hypothesized that the strength of the connectivity between some regions would differ MI performance between the two groups. In addition, we analyze the effective connectivity of the resting basic network and examine its association with MI performance. In recent years, the default mode network of resting-state has been increasingly used to investigate cognitive performance [30-31]. This may be advantageous to observe new changes in investigating the variability of MI performance excluding areas with MI-related characteristics. These findings would show the possibility that effective connectivity would be used as an indicator of performance to predict MI performance. Study on this topic may be instructive for the establishment of enhanced training strategies for subjects who exhibit poor performance on these tasks.

## II. METHODS

### A. EEG Dataset

The datasets used for this work were the GigaDB provided by Gigascience [32]. The associated EEG data were acquired from 52 subjects. Fig. 1 shows the design of the whole experiment. This work includes each work before and after the BCI task. Before the MI task, a questionnaire about the information and status of subjects, and 1-min resting-state EEG data with opened-eyes were included. Then, MI task of five runs was performed. For each run, 100 or 120 trials were conducted for the MI of left and right-hand movement, respectively.

### B. MI Performancey and Categorization

In this work, the overall data analysis was performed using MATLAB R2019b. The MI performance of each subject was quantified as MI performance using code provided in GigaDB. All trials for each subject were filtered spectrally 8-30 Hz. Common spatial patterns were used to extract the MI-related EEG features. Then, the extracted features classify the left and right-hand MI conditions using Fisher's linear discriminant analysis. In this work, we have used cross-validation for yield a statistically reasonable BCI performance. Every trial was grouped into ten sets. These ten sets separated into 7 training and 3 testing sets. Finally, the MI classification accuracy of each subject was obtained [33]. We grouped all subjects into two groups to investigate common feature differences between subjects with high and low MI performance.

### C. Effective Connectivity in Resting-State EEG

Resting-state EEG of 1-min was pre-processed using the EEGLAB toolbox [34]. The data was band-pass filtered to 4-45 Hz and epochs between 0 and 1000 msec. Then, the whole epochs were averaged. Effective connectivity was estimated quantitatively using DCM. This method was analyzed using the SPM12 toolbox [35].

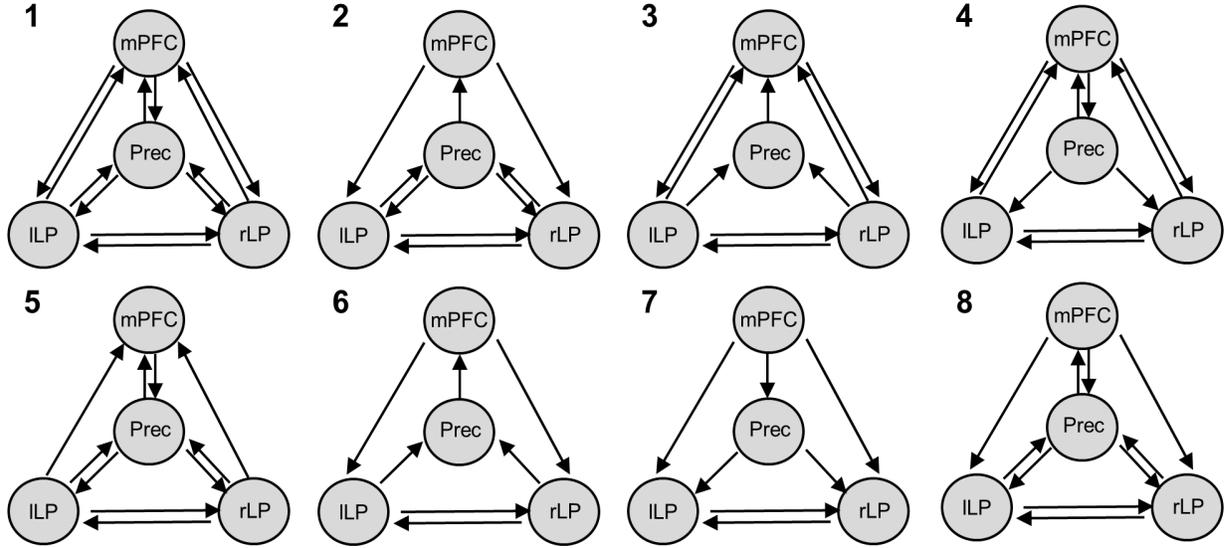

Figure 2. The models specification of regions of interest (ROI) in default mode network. l/rLP = left and right lateral parietal area, Prec = posterior cingulate/precuneus, mPFC = medial prefrontal cortex.

First, we specified an anatomical location for the ROI. The ROI specified as the most used default mode network, the default mode network includes four nodes. The nodes are as follows: left and right lateral parietal area (l/rLP), posterior cingulate/precuneus (Prec) and medial prefrontal cortex (mPFC) [36]. Table I shows the montreal neurological institute coordinates of the default mode network. DCM models for effective connectivity in resting-state EEG have constructed the eight models from anatomical and structural imaging and computational modeling. Fig. 2 shows the eight models specified. Then, parameters for connectivity strength were estimates from each model. We applied Bayesian model selection (BMS) from 8 models for 52 subjects using fixed-effects analysis to determine the most likely model given the data.

### D. Statistical Analysis

To the reliability of the proposed predictors based on effective connectivity and investigation of the change for connectivity strength between high and low BCI performance groups, we performed two statistical analyses. First, the two sample t-test was performed to investigate the difference in connectivity strength of connection between high and low MI performance groups. Second, we analyzed the Pearson's correlation between MI classification accuracy and connection for effective connectivity of all subjects. All significance level of statistical analysis is $\alpha = 0.05$.

## III. RESULTS

### A. MI Performance

MI classification accuracy was calculated for all subjects. The average MI classification accuracy of two class for left and right hand movement from all subjects was 65% ± 9.2%. We divided each of the 26 subjects into two groups from all subjects. For each group, a mean MI classification accuracy of 72 ± 8.9% in the high group and low group obtained mean MI classification accuracy of 59 ± 2.4% ($t = 7.127$, $p < 0.001$).

### B. BMS Results

BMS for each of the eight models from 52 subjects were analyzed. As a result, the model 7 was chosen as the most likely mode for the given data. Fig. 3 shows the results of BMS

TABLE I. Montreal Neurological Institute (MNI) Coordinates for Location Information of ROI

| | MNI coordinates | | |
|---|---|---|---|
| ROI[a] | X | Y | Z |
| lLP | -46 | -66 | 30 |
| rLP | 49 | -63 | 33 |
| Prec | 0 | -58 | 0 |
| mPFC | -1 | 54 | 27 |

a. Region of interest.

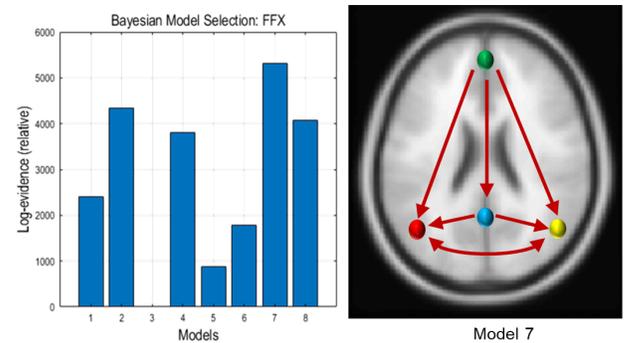

Figure 3. Results of Bayesian model selection. Left: The model 7 selected as a result of log-evidence values for eight models using Bayesian model selection. Right: selected model 7. Nodes; green: mPFC (medial prefrontal cortex), red: lLP (left lateral parietal area), yellow: rLP (right lateral parietal area), and blue: prec (posterior cingulate/precuneus).

in DCM and selected model 7. This model includes total seven connections and each connection signify connectivity strength.

*C. Relationship of Connectivity Strength between High and Low MI Groups*

We investigate the connectivity strength of the estimated parameters between high and low groups. As a result, there was a significant difference in connectivity strength from lLP to rLP connection between high and low groups ($t = -2.03$, $p = 0.048$). Specifically, connection of from lLP to rLP showed stronger connection strength in the low group (mean connectivity strength: $0.70 \pm 0.15$) compared to in the high group (mean connectivity strength: $0.62 \pm 0.09$). Table II showed the t-test results for effective connectivity in resting-state between high and low groups. In addition, correlation between effective connectivity of all subjects and MI classification accuracies was analyzed. As a result, the connectivity from lLP to rLP, which showed a significant difference in connections between the two groups, was significantly correlated with the MI classification accuracy ($r = -0.37$, $p = 0.007$). However, there was no significant correlation with the MI classification accuracy at other connections. Table III showed the statistical results of the correlation with the MI classification accuracies. Fig. 4 showed the correlation with MI classification accuracies for connection from lLP to rLP.

## IV. DISCUSSION AND CONCLUSION

We investigated the difference between the high and low groups and the correlation with MI classification accuracy from the effective connectivity of all subjects. In summary, we found a significant correlation between MI-BCI performance and effective connectivity in resting-state EEG. Statistical analysis of the effective connectivity between high and low

TABLE II.  Statistical Results for Comparison of Connectivity Strength between High and Low Groups

| From | To | *t*-value | *p*-value |
|---|---|---|---|
| rLP[a] | lLP | 0.207 | 0.837 |
| Prec[b] | | -0.657 | 0.514 |
| mPFC[c] | | -1.560 | 0.125 |
| lLP | rLP | -2.030 | **0.048** |
| Prec | | -2.009 | 0.050 |
| mPFC | | -0.132 | 0.896 |
| mPFC | Prec | -0.146 | 0.885 |

a. Left and right lateral parietal area (l/rLP), b. Posterior cingulate/precuneus (Prec), c. Medial prefrontal cortex (mPFC).

TABLE III.  Correlation with MI Performance for Connecitivity Strength in Resting-state EEG from All Subjects

| From | To | rho | *p*-value |
|---|---|---|---|
| rLP[a] | lLP | 0.110 | 0.436 |
| Prec[b] | | 0.059 | 0.680 |
| mPFC[c] | | -0.138 | 0.329 |
| lLP | rLP | -0.367 | **0.007** |
| Prec | | -0.180 | 0.201 |
| mPFC | | 0.040 | 0.780 |
| mPFC | Prec | -0.171 | 0.226 |

a. Left and right lateral parietal area (l/rLP), b. Posterior cingulate/precuneus (Prec), c. Medial prefrontal cortex (mPFC).

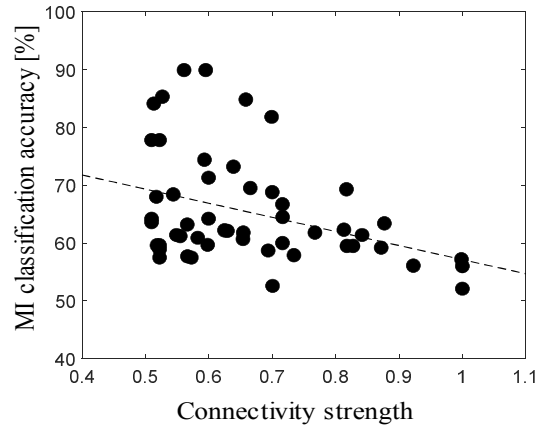

Figure 4.  Correlation between MI performance and connection from rLP to lLP for all subjects. lLP (left lateral parietal area), rLP (right lateral parietal area).

groups archives consistent results that subjects with better MI classification accuracy have stronger connectivity strength for connection from lLP to rLP. Also, this connection correlated with MI classification accuracy of all subjects. The connection shows a positive correlation.

Performance variation of inter-subjects is considered as one of the biggest obstacles for MI-BCI. We have investigated the MI classification accuracy from all subjects to analyze the difference in MI performances of two groups. In Giga database, MI-BCI performance between two groups shows a distinct difference. Müller et al. [37] were reported that the upper confidence limits of chance with α = 5% were 70% (classification accuracy) in a 2-class problem. In our results, all subjects obtain classification accuracy of chance level but most subjects had not above 70%.

We observed the strong correlation between MI performance and resting-state default mode network. The resting-state default mode network is associated with motor learning [38]. In specific, these network including medial temporal and lateral parietal regions increases after motor imagery [39]. In this regard, the activation of resting-state default mode network before MI is considered to have a direct and indirect effect on MI.

Our study has some limitations. First, we need a DCM analysis that adds MI-related ROI for effective connectivity in this study. We set up the most fundamental default mode network in DCM analysis, but we expect more significant predictors if we add MI-related ROI and frontal regions. Second, there is a need to investigate common predictors of performance variation in the various datasets. Sannellli et al. [9] emphasized the importance of analyzing common causes for performance variation of large scale subjects.

In conclusion, our results show that the connection between the left and right lateral parietal in effective connectivity from resting-state EEG is highly related to the MI classification accuracy. These works could that proposed predictors are helpful in better understanding the mechanisms of the MI-BCI and may help to find new strategies for improving MI-BCI performance.

## REFERENCES

[1] M. Lee, C.-H. Park, C.-H. Im, J.-H. Kim, G.-H. Kwon, L. Kim, W. H. Chang, and Y.-H. Kim, "Motor imagery learning across a sequence of trials in stroke patients," *Restor. Neurol. Neurosci.*, vol. 34, pp. 635-645, Aug. 2016.

[2] S. Halder, D. Agorastos, R. Veit, E. M. Hammer, S. Lee, B. Varkuti, M. Bogdan, W. Rosenstiel, N. Birbaumer, and A. Kübler, "Neural mechanisms of brain-computer interface control," *Neuroimage*, vol. 55, pp. 1779-1790, Apr. 2011.

[3] A. Kübler, N. Neumann, "Brain–computer interfaces–the key for the conscious brain locked into a paralyzed body," *Prog. Brain Res.*, vol. 150, pp. 513-525, 2005.

[4] N.-S. Kwak, K.-R. Müller, and S.-W. Lee, "A convolutional neural network for steady state visual evoked potential classification under ambulatory environment," *PLoS One*, vol. 12, pp. e0172578, Feb. 2017.

[5] H.-I. Suk and S.-W. Lee, "Subject and class specific frequency bands selection for multiclass motor imagery classification," *Int. J. Imaging Syst. Technol.*, vol. 21, pp. 123-130, May 2011.

[6] J.-H. Kim, F. Bießmann, and S.-W. Lee, "Decoding three-dimensional trajectory of executed and imagined arm movements from electroencephalogram signals," *IEEE Trans. Neural Syst. Rehabil. Eng.*, vol. 23, pp. 867-876, Sep. 2015.

[7] S.-K. Yeom, S. Fazli, K.-R. Müller, and S.-W. Lee, "An efficient ERP-based brain-computer interface using random set presentation and face familiarity," *PLoS One*, vol. 9, pp. e111157, Nov. 2014.

[8] S. Rimbert, N. Gayraud, L. Bougrain, M. Clerc, and S. Fleck, "Can a subjective questionnaire be used as brain-computer interface performance predictor?," *Front. Hum. Neurosci.*, vol. 12, pp. 529, Jan. 2018.

[9] M.-H. Lee, O.-Y. Kwon, Y.-J. Kim, H.-K. Kim, Y.-E. Lee, J. Williamson, S. Fazli, and S.-W. Lee, "EEG dataset and OpenBMI toolbox for three BCI paradigms: an investigation into BCI illiteracy," *GigaScience*, vol. 8, pp. giz 002, Jan. 2019.

[10] C. Sannelli, C. Vidaurre, K.-R. Müller, and B. Blankertz, "A large scale screening study with a SMR-based BCI: Categorization of BCI users and differences in their SMR activity," *PLoS One*, vol. 14, pp. e0207351, Jan. 2019.

[11] Y. Chen, A. D. Atnafu, I. Schlattner, W. T. Weldtsadik, M. C. Roh, H. J. Kim, S.-W. Lee, B. Blankertz, and S. Fazli, "A high-security EEG-based login system with RSVP stimuli and dry electrodes," *IEEE Trans. Inf. Forensic Secur.*, vol. 11, pp. 2635-2647, Dec. 2016.

[12] A. Myrden, and T. Chau, "Effects of user mental state on EEG-BCI performance," *Front. Hum. Neurosci.*, vol. 9, pp. 000308, Jan. 2015.

[13] C. Jeunet, B. N'Kaoua, S. Subramanian, M. Hachet, and F. Lotte, "Predicting mental imagery-based BCI performance from personality, cognitive profile and neurophysiological patterns," *PLoS One*, vol. 10, pp. e0143962, Dec. 2015.

[14] M. Ahn, H. Cho, S. Ahn, and S. C. Jun, "User's self-prediction of performance in motor imagery brain–computer interface," *Front. Hum. Neurosci.*, vol. 12, pp. 00059, Feb. 2018.

[15] B. Blankertz, C. Sannelli, S. Halder, E. M. Hammer, A. K.-R. Müller, G. Curio, and T. Dickhaus, "Neurophysiological predictor of SMR-based BCI performance," *Neuroimage*, vol. 51, pp. 1303-1309, July 2010.

[16] M. Ahn, H. Cho, S. Ahn, and S. C. Jun, "High theta and low alpha powers may be indicative of BCI-illiteracy in motor imagery," *PLoS One*, vol. 8, pp. e80886, Nov. 2013.

[17] R. Zhang, D. Yao, P. A. Valdés-Sosa, F. Li, P. Li, T. Zhang, T. Ma, Y. Li, and P. Xu, "Efficient resting-state EEG network facilitates motor imagery performance," *J. Neural Eng.*, vol. 12, pp. e80886, Nov. 2015.

[18] X. Zhu, H.-I. Suk, S.-W. Lee, and D. Shen, "Canonical feature selection for joint regression and multi-class identification in Alzheimer's disease diagnosis," *Brain Imaging Behav.*, vol. 10, pp. 818-828, Sep. 2016.

[19] H. H. Bülthoff, S.-W. Lee, T. A. Poggio, and C. Wallraven, "Biologically motivated computer vision," *NY:Springer-Verlag*, 2003.

[20] M. Kim, G. Wu, Q. Wang, S.-W. Lee, and D. Shen, "Improved image registration by sparse patch-based deformation estimation," *Neuroimage*, vol. 105, pp. 257-268, Jan. 2015.

[21] R. Zhang, X. Li, Y. Wang, B. Liu, L. Shi, M. Chen, L. Zhang, and Y. Hu, "Using brain network features to increase the classification accuracy of MI-BCI inefficiency subject," *IEEE Access*, vol. 7, pp. 74490-74499, May 2019.

[22] M. Lee, R. D. Sanders, S.-K. Yeom, D.-O. Won, K.-S. Seo, H. J. Kim, G. Tononi, and S.-W. Lee, "Network properties in transitions of consciousness during propofol-induced sedation," *Sci. Rep.*, vol. 7, pp. 16791, Dec. 2017.

[23] K. J. Friston, "Functional and effective connectivity in neuroimaging: A synthesis," *Hum. Brain Mapp.*, vol. 2, pp. 56-78, May 1994.

[24] M. Lee, B. Baird, O. Gosseries, J. O. Nieminen, M. Boly, B. R. Postle, G. Tononi, and S.-W. Lee, "Connectivity differences between consciousness and unconsciousness in non-rapid eye movement sleep: a TMS–EEG study," *Sci. Rep.*, vol. 9, pp. 5175, Mar. 2019.

[25] X. Ding and S.-W. Lee, "Changes of .functional and effective connectivity in smoking replenishment on deprived heavy smokers: a resting-state fMRI study," *PLoS One*, vol. 8, pp. e59331, Mar. 2013.

[26] A. K. Rehme, S. B. Eickhoff, and C. Grefkes, "State-dependent differences between functional and effective connectivity of the human cortical motor system," *Neuroimage*, vol. 67, pp. 237-246, Feb. 2013.

[27] H. Huang, X. Liu, Y. Jin, S.-W. Lee, C.-Y. Wee, and D. Shen, "Enhancing the representation of functional connectivity networks by fusing multi view information for autism spectrum disorder diagnosis," *Hum. Brain Mapp.*, vol. 40, pp. 833-854, Oct. 2018.

[28] O. David, S. J. Kiebel, L. M. Harrison, J. Mattout, J. M. Kilner, and K. J. Friston, "Dynamic causal modeling of evoked responses in EEG and MEG," *Neuroimage*, vol. 30, pp. 1255-1272, May 2006.

[29] Y. K. Kim, E. Park, A. Lee, C.-H. Im, and Y.-H. Kim, "Change in network connectivity during motor imagery and excution," *PLoS One*, vol. 13, pp. e0190715, Jan. 2018.

[30] M. Alavash, P. Doebler, H. Holling, C. M. Thiel, and C. Gießinga, "Is functional integration of resting state brain networks an unspecific biomarker for working memory performance?," *Neuroimage*, vol. 108, pp. 182-193, Mar. 2015.

[31] S. Markett, M. Reuter, C. Montag, G. Voigt, B. Lachmann, S. Rudorf, C. E. Elger, and B. Weber, "Assessing the function of the fronto-parietal attention network: insights from resting-state fMRI and the attentional network test," *Hum. Brain Mapp.*, vol. 35, pp. 1700-1709, Apr. 2014.

[32] H. Cho, M. Ahn, S. Ahn, M. Kwon, and S. C. Jun, "EEG datasets for motor imagery brain-computer interface," *GigaScience*, vol. 6, pp. 1-8, July 2017.

[33] B. Blankertz, R. Tomioka, S. Lemm, M. Kawanabe, and K.-R. Müller, "Optimizing spatial filters for robust EEG single-trial analysis," *IEEE Signal Process. Mag.*, vol. 25, pp. 41-56, Jan. 2008.

[34] A. Delorme and S. Makeig, "EEGLAB: an open source toolbox for analysis of single-trial EEG dynamics including independent component analysis," *J. Neurosci. Methods*, vol. 134, pp. 9-21, March 2004.

[35] M. Brett, J.-L. Anton, R. Valabregue, and J.-B. Poline, "Region of interest analysis using an SPM toolbox," *Proc. 8th Int. Conf. on Functional Mapping of the Human Brain*, vol. 16, pp. 497, June 2002.

[36] F. V. de Steen, H. Almgren, A. Razi, K. Friston, and D. Marinazzo, "Dynamic causal modelling of fluctuating connectivity in resting-state EEG," *Neuroimage*, vol. 189, pp. 476-484, Apr. 2019.

[37] G. R. Müller, R. Scherer, C. Brunner, R. Leeb, and G. Pfurtscheller, "Better than random? A closer look on BCI results," *Int. J. Bioelectromagn*, vol. 10, pp. 52-55, Jan. 2008.

[38] L. Ma, S. Narayana, D. A. Robin, P. T. Fox, and J. Xiong, "Changes occur in resting state network of motor system during 4 weeks of motor skill learning," *Neuroimage*, vol. 58, pp. 226-233. Sep. 2011.

[39] R. Ge, H. Zhang, L. Yao, and Z. Long, "Motor imagery learning induced changes in functional connectivity of the default mode network," *IEEE Trans. Neural Syst. Rehabil. Eng.*, vol. 23, pp. 138-148, Jan. 2014.